\documentclass{PoS}
\usepackage[utf8x]{inputenc}
\usepackage{amsmath}
\usepackage{mathtools}
\usepackage{nicefrac}
\usepackage{bm}
\usepackage{float}
\title{Flux tubes and coherence length in the $SU(3)$ vacuum}

\ShortTitle{Flux tubes and coherence length in the $SU(3)$ vacuum}

\author{Paolo Cea\\
        Dipartimento di Fisica dell’Università di Bari \\and INFN - Sezione di Bari, I-70126 Bari, Italy\\
        E-mail: \email{paolo.cea@ba.infn.it}}
\author{Leonardo Cosmai\\
        INFN - Sezione di Bari, I-70126 Bari, Italy\\
        E-mail: \email{leonardo.cosmai@ba.infn.it}}
\author{\speaker{Francesca Cuteri}\\
        Dipartimento di Fisica dell'Università della Calabria \\and INFN - Gruppo collegato di Cosenza, I-87036 Arcavacata di Rende, Cosenza, Italy\\
        E-mail: \email{francesca.cuteri@fis.unical.it}}
\author{Alessandro Papa\\
        Dipartimento di Fisica dell'Università della Calabria \\and INFN - Gruppo collegato di Cosenza, I-87036 Arcavacata di Rende, Cosenza, Italy\\
        E-mail: \email{alessandro.papa@fis.unical.it}}

\abstract{An estimate of the London penetration and coherence lengths in the vacuum of the $SU(3)$ pure gauge theory is given downstream an analysis of the transverse profile of the chromoelectric flux tubes. Within ordinary superconductivity, a simple variational model for the magnitude of the normalized order parameter of an isolated vortex produces an analytic expression for magnetic field and supercurrent density. In the picture of $SU(3)$ vacuum as dual superconductor, this expression provides us with the function that fits the chromoelectric field data. The smearing procedure is used in order to reduce noise.}

\FullConference{31st International Symposium on Lattice Field Theory - LATTICE 2013\\
		July 29 - August 3, 2013\\
		Mainz, Germany}

\begin{document}

\section{Introduction}
Since the absence of colored states in the spectrum of QCD and the inability to detect free quarks have not been explained from first principles yet, the color confinement phenomenon still represents an open problem~\cite{1}.
In the picture of the dual superconductor model for color confinement~\cite{2}, a bound state of a quark-antiquark pair in the QCD vacuum, at low temperature and density, may be thought of as a chromoelectric flux tube connecting the two charges.
Through the 't Hooft construction, a physical analogy is traced between the QCD vacuum and an ordinary electrical superconductor: the former behaves like a coherent state of color magnetic monopoles in almost the same way as the latter behaves as a coherent state of Cooper pairs~\cite{Ripka:2003vv}. With the QCD vacuum behaving like a dual superconductor, the observation of tube-like structures is an indication of the presence of a linear potential proving, from a phenomenological point of view, the color confinement phenomenon~\cite{4,5,DiGiacomo:1990hc}.
Chromoelectric flux tubes, in this scenario, represent the dual analog of Abrikosov vortices in the ordinary superconductivity~\cite{Abrikosov:1956sx,clem1975simple,PhysRev.166.447,Charbonneau:2007db}. The effect accounting for their presence in the vacuum is the dual version of the Meissner effect.

Previous studies about the $SU(3)$ confining vacuum have revealed that color flux tubes are almost completely formed by the longitudinal chromoelectric field $E_{l}$, which is constant along the line joining a static quark-antiquark pair, while rapidly decreasing in the transverse direction $x_{t}$.
In this last direction, $E_{l}\left(x_{t}\right)$ has been fitted by means of functions coming from the theory of ordinary superconductivity and being rewritten according to the dual analogy~\cite{4,5}.

In the present work, in view of extending the studies about flux tubes in $SU(3)$ vacuum to the case of finite temperature, as a preliminary step, a correlator between two Polyakov loops is implemented on the lattice. One of the Polyakov loops is connected to a plaquette, to measure, as first, the
field at zero temperature. Before discussing the results of numeric measurements and comparing them with the ones produced when the correlator of a Wilson loop with a plaquette was considered, the two possible fitting functions for $E_{l}(x_{t})$ are traced back to both the London model and Ginzburg-Landau theory for the ordinary superconductivity~\cite{Abrikosov:1956sx,clem1975simple,PhysRev.166.447}.

\section{Fit functions}

Two main proposals for the shape of $E_{l}\left(x_{t}\right)$ will be considered. The first one \cite{4}, usually said to be derived from the London model for the ordinary superconductivity, is
\begin{equation}
 E_{l}\left(x_{t}\right)=\frac{\Phi}{2\pi}\mu^{2}K_{0}\left(\mu x_{t}\right),\qquad\qquad x_{t}>0,
\label{eq:fitfunct1}
\end{equation}
while the second one~\cite{5} is
\begin{equation}
 E_{l}\left(x_{t}\right)=\frac{\Phi}{2\pi}\frac{\mu^{2}}{\alpha}\frac{K_{0}\left[\left(\mu^{2}x_{t}^{2}+\alpha^{2}\right)^{\nicefrac{1}{2}}\right]}{K_{1}\left[\alpha\right]}.
\label{eq:fitfunct2}
\end{equation}
In the expressions above, $K_{0}$ and $K_{1}$ are the modified Bessel functions of order zero and one respectively, $\Phi$ is the external flux, $\lambda=\nicefrac{1}{\mu}$ is the London penetration length and $\alpha=\nicefrac{\xi_{\nu}}{\lambda}$ is a quantity which is tied to $\xi_{\nu}$ (a variational core radius parameter of the order of the coherence length $\xi$ of the magnetic monopole condensate), hence to the Ginzburg-Landau parameter $\kappa$, distinguishing type I and type II superconductors, by means of the relation
\begin{equation}
 \kappa=\frac{\sqrt{2}}{\alpha}\left[1-\frac{K_{0}^{2}\left(\alpha\right)}{K_{1}^{2}\left(\alpha\right)}\right]^{\nicefrac{1}{2}}.
\end{equation}
The main limitations concerning the use of the function (\ref{eq:fitfunct1}) are that it diverges on the axis of the vortex line, providing realistic values only far away from the vortex core ($x_{t}>0$), and that it gives an adequate description of the transverse structure of the flux tube only if $\kappa=\nicefrac{\lambda}{\xi}\gg1$, that is only in the Shubnikov phase, far enough from the upper critical field.
Conversely, the function (\ref{eq:fitfunct2}) yields realistic values for the field also in the vortex core vicinity and it reduces to (\ref{eq:fitfunct1}) far outside the vortex core.
In what follows it will be shown that the choice of the theory from which one derives the expressions for $E_{l}\left(x_{t}\right)$ is just a matter of convenience: both functions can be derived, under proper assumptions and dual transformation, from both London and Ginzburg-Landau theories.

\section{Fit functions from London Model}
The phase diagram for type II superconductors shows the existence of two critical fields $B_{c1}$ (lower) and $B_{c2}$ (upper). In between $B_{c1}$ and $B_{c2}$, the superconductor is in the so-called mixed state. The lower critical field is the one at which the first vortex line appears in the superconductor. Slightly above $B_{c1}$ only a few vortex lines occur, separated by a distance much greater than $\lambda$, such that they can be treated as isolated. This allows the use of the London model with the assumption that the diameter $\left(2\xi\right)$ of the normal vortex core is very small compared to $\lambda$, i.e. $\kappa\gg 1$.
In the limit $\kappa\rightarrow\infty$, the vortices become line singularities and can be described by the modified London equation
\begin{equation}
\mu_{0}\lambda^{2}\boldsymbol\nabla\times\boldsymbol J+\boldsymbol B=\Phi_{0}\delta\left(\boldsymbol r-\boldsymbol r_{0}\right)\hat{\boldsymbol{z}},\label{eq:lond1}
\end{equation}
where $\Phi_{0}$ is the flux quantum of the vortex line, while $\lambda=\nicefrac{1}{\mu}$ is the London penetration depth. From (\ref{eq:lond1}) one obtains $\lambda^{2}\boldsymbol\nabla^{2}\boldsymbol B -\boldsymbol B=-\Phi_{0}\delta\left(\boldsymbol{r}-\boldsymbol{r_{0}}\right)\hat{\boldsymbol{z}}$, which is solved by
\begin{equation}
\boldsymbol B\left(r\right)=\frac{\Phi_{0}}{2\pi}\mu^{2}K_{0}\left(\mu r\right)\hat{\boldsymbol{z}}.
\end{equation}
Then, (\ref{eq:fitfunct1}) is obtained from this ordinary superconductivity result by use of the dual analogy.

Few extra assumptions are needed if one is to obtain the expression (\ref{eq:fitfunct2}). One tries to avoid the infinity at the vortex core by cutting off the magnetic field, applied on the $z$-axis of an infinitely long cylindrical superconductor, within a certain critical radius $r_{0}$~\cite{PhysRev.166.447}. In the region $r<r_{0}$, the field $B$ is assumed to be uniform and equal to $B\left(\nicefrac{r_{0}}{\lambda}\right)$, while for $r>r_{0}$ the field satisfies London equation $\boldsymbol\nabla^{2}\boldsymbol B =\frac{\boldsymbol B}{\lambda^{2}}$ that, if variations of $B$ in the $\theta$ and $z$ directions are ignored, is solved by
\begin{equation}
B\left(\frac{r}{\lambda}\right)=CI_{0}\left(\frac{r}{\lambda}\right)+DK_{0}\left(\frac{r}{\lambda}\right).
\end{equation}
The cut-off allows us to retain the $K$-function solutions, while the $I$-function solutions increasing monotonically with $r$ are excluded, $C=0$, for the case of an isolated fluxoid. Moreover, the London fluxoid quantization condition, 
\begin{equation}
 \iint \boldsymbol B \cdot d\boldsymbol s +\frac{4\pi\lambda^{2}}{c}\ointop_{R}\boldsymbol J \cdot d\boldsymbol l =n\Phi_{0},\label{eq:quantization}
\end{equation}
holds since, inside $r_{0}$, the sample is in the normal state. By taking the contour $R$ at infinity, where $\boldsymbol J=\boldsymbol 0$, the integral is evaluated and $D$ determined to give
\begin{equation}
B\left(\frac{r}{\lambda}\right)=\frac{n\Phi_{0}-\pi r_{0}^{2}B\left(\nicefrac{r_{0}}{\lambda}\right)}{2\pi\lambda r_{0}}\frac{K_{0}\left(\nicefrac{r}{\lambda}\right)}{K_{1}\left(\nicefrac{r_{0}}{\lambda}\right)},\qquad r\ge r_{0}.
\end{equation}
Now, the easiest way to smoothly extend this solution also in the region $r<r_{0}$, where the cut-off is active, is to make the substitution $r\rightarrow\sqrt{r^{2}+r_{0}^{2}}$. Then, by interpreting $r_{0}$ as $\xi_{\nu}$, and invoking the dual analogy again, (\ref{eq:fitfunct2}) is obtained.

\section{Fit functions from Ginzburg-Landau theory}
In the context of Ginzburg-Landau theory for ordinary superconductivity, vortex line solutions come from solving the Ginzburg-Landau equations in cylindrical coordinates, with the magnetic flux line centered on the $z$-axis. The fact is that one has to do with coupled non-linear differential equations.
However, if one choices an ansatz for the normalized order parameter $\Psi$, which takes into account the depression of $\Psi$ to zero on the axis and has the correct limiting behavior (reproducing London results for $r\gg0$), $\boldsymbol B$ can be obtained by solving the second Ginzburg-Landau equation.
Similarly, an ansatz for the potential vector can be introduced, with reasonable boundary conditions. One can obtain (\ref{eq:fitfunct1}) by assuming~\cite{Charbonneau:2007db}
\begin{equation}
\Psi=\sqrt{\frac{v}{u}}\rho\left(r\right)\exp\left(i\varphi\left(r\right)\right),
\end{equation}
and the ansatz
\begin{equation}
 \boldsymbol{A}=\frac{\hbar c}{q}\frac{a\left(r\right)}{r}\hat{\boldsymbol{\varphi}}
\end{equation}
for the vector potential, inspired by the fact that, far from the vortex core, where $\boldsymbol J =\boldsymbol 0$, the second Ginzburg-Landau equation requires $\boldsymbol{A}=\frac{\hbar c}{q}\boldsymbol{\nabla}\varphi\left(r\right)$.
Clearly, the functions $\rho\left(r\right)$ and $a\left(r\right)$ in the expressions for $\Psi$ and $\boldsymbol A$ must be such that $\rho\left(r\right),\, a\left(r\right)\rightarrow1$ as $r\rightarrow\infty$, while $\rho\left(r\right),\, a\left(r\right)\rightarrow0$ as $r\rightarrow 0$. Hence, far away from the vortex axis, the expansions
\begin{equation}
\rho\left(r\right)=1+\sigma\left(r\right),\quad a\left(r\right)=1+r\alpha\left(r\right);\quad\sigma\left(r\right),\alpha\left(r\right)\rightarrow 0\,\quad\mathrm{as}\ \quad r\rightarrow\infty,
\end{equation}
are admitted and the second Ginzburg-Landau equation reads
\begin{equation}
\frac{d^{2}\alpha\left(r\right)}{dr^{2}}+\frac{1}{r}\frac{d\alpha\left(r\right)}{dr}-\frac{\alpha\left(r\right)}{r^{2}}=\frac{4\pi q^{2}v}{mc^{2}u}\alpha\left(r\right)\left(1+\sigma\left(r\right)\right)^{2}.
\end{equation}
Linearization, by taking $r\rightarrow\infty$ and keeping only terms linear in $\alpha\left(r\right)$ and $\sigma\left(r\right)$, yields a modified, homogeneous Bessel equation of the first order
\begin{equation}
\frac{d^{2}\alpha\left(r\right)}{dr^{2}}+\frac{1}{r}\frac{d\alpha\left(r\right)}{dr}-\left(\frac{1}{r^{2}}+\frac{1}{\lambda^{2}}\right)\alpha\left(r\right)=0,\qquad\lambda=\sqrt{\frac{mc^{2}u}{4\pi q^{2}v}}.
\end{equation}
The solution is the modified Bessel function of the second kind $\alpha\left(r\right)=\nicefrac{K_{1}\left(\nicefrac{r}{\lambda}\right)}{\lambda}$, from which
\begin{equation}
\boldsymbol{A}=\frac{\hbar c}{q}K_{1}\left(\frac{r}{\lambda}\right)\hat{\boldsymbol{\varphi}}.
\end{equation}
Applying the curl gives us $\boldsymbol{B}=\frac{\Phi_{0}}{2\pi}\mu^{2}K_{0}\left(\mu r\right)\hat{\boldsymbol{z}}$ that, through the dual analogy, becomes (\ref{eq:fitfunct1}).

To obtain (\ref{eq:fitfunct2}), according to the method presented in~\cite{clem1975simple} one has to use, as a variational trial function, $f=\nicefrac{r}{R}$, with $R=\left(r^{2}+\xi_{\nu}^{2}\right)^{\nicefrac{1}{2}}$, where $r$ is the radial coordinate and $\xi_{\nu}$ is a variational core radius parameter.
With $\boldsymbol{b}=b_{z}\left(r\right) \hat{\boldsymbol z}$ , $\boldsymbol{j}=j_{\varphi}\left(r\right)\hat{\boldsymbol{\varphi}}$, $\boldsymbol{a}=a_{\varphi}\left(r\right)\hat{\boldsymbol{\varphi}}$ in the Coulomb gauge, and with
\begin{equation}
\Psi\left(r\right)=f\left(r\right)\exp\left(-i\varphi\right),\qquad f\left(r\right)=\nicefrac{r}{\left(r^{2}+\xi_{\nu}^{2}\right)^{\nicefrac{1}{2}}},
\end{equation}
the second Ginzburg-Landau equation reads
\begin{equation}
\boldsymbol{j}=-\frac{c}{4\pi\lambda^{2}}\left[a_{\varphi}-\frac{\Phi_{0}}{2\pi r}\right]f^{2}\hat{\boldsymbol{\varphi}}.
\end{equation}
The corresponding differential equation for $a_{\varphi}$ and its solution are
\begin{equation}
\frac{d}{dr}\left[\frac{1}{r}\frac{d}{dr}\left(\rho a_{\varphi}\right)\right]-\frac{f^{2}}{\lambda^{2}}a_{\varphi}=-\frac{\Phi_{0}f^{2}}{2\pi\lambda^{2}r},\qquad a_{\varphi}=\frac{\Phi_{0}}{2\pi r}\left[1-\frac{RK_{1}\left(\nicefrac{R}{\lambda}\right)}{\xi_{\nu}K_{1}\left(\nicefrac{\xi_{\nu}}{\lambda}\right)}\right].
\end{equation}
One can, then, compute the magnetic flux and this leads, finally, to (~\ref{eq:fitfunct2}).

\begin{figure}[tb] 
  \centering
    \includegraphics[width=0.26\textwidth]{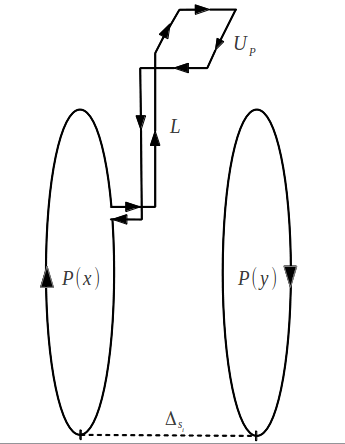} 
    \includegraphics[width=0.50\linewidth]{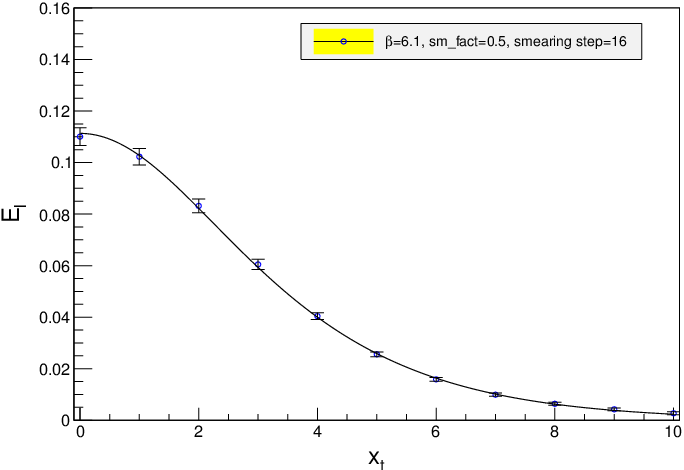} 
  \caption{(Left) The connected correlator (\protect\ref{eq:rhopconn}) between the plaquette $U_{P}$ and the Polyakov loops (subtraction in $\rho_{P}^{conn}$ not explicitly drawn). (Right) $E_{l}\left(x_{t}\right)$ versus $x_{t}$ at $\beta = 6.1$ after 16 smearing steps.}\label{loop&field}
\end{figure}

\section{Color field measure on the lattice}

\begin{figure}[tb] 
  \centering
   \includegraphics[width=0.495\linewidth]{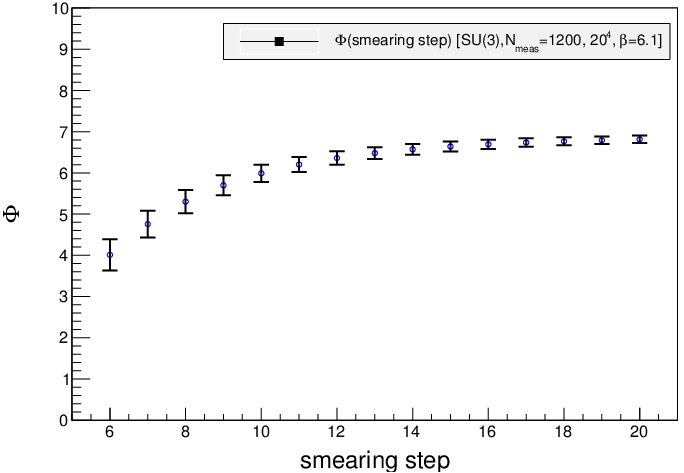}
   \includegraphics[width=0.495\linewidth]{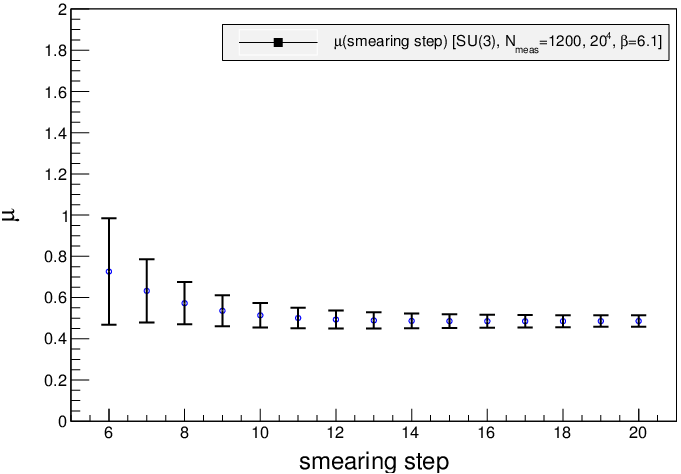}
   \includegraphics[width=0.495\linewidth]{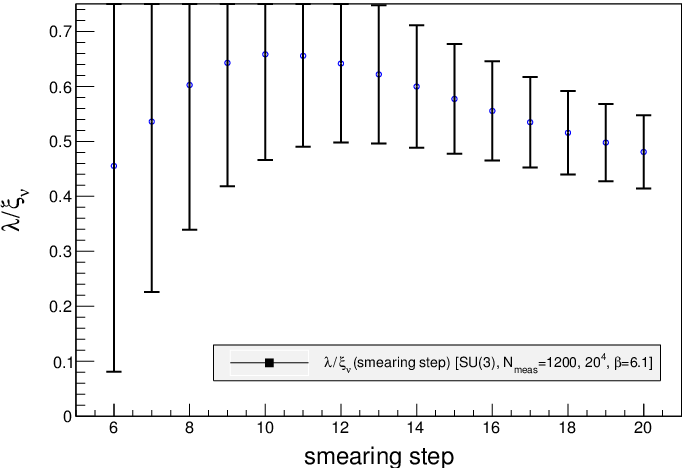}
   \includegraphics[width=0.495\linewidth]{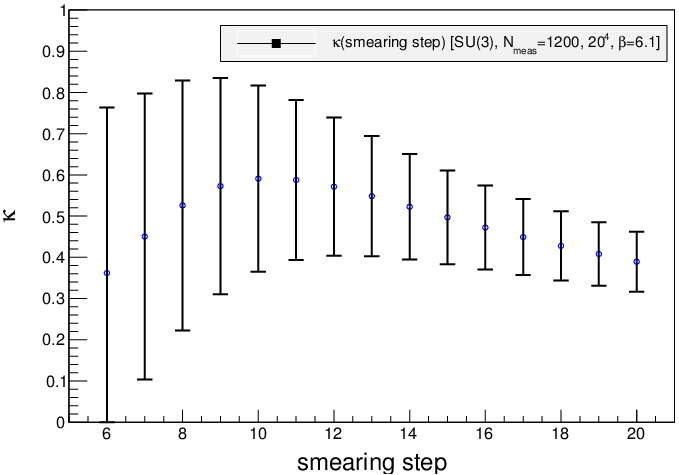}
  \caption{(Up left) The inverse of the penetration length $\mu$ at $\beta=6.1$ versus the smearing step. (Up right) The same for the amplitude of the longitudinal chromoelectric field $\Phi$. (Down left) The same for $\nicefrac{\lambda}{\xi_{\nu}}$. (Down right) The same for the Ginzburg-Landau parameter $\kappa$.}\label{param}
\end{figure}

In order to measure the color field for a static quark-antiquark pair in the $SU(3)$ vacuum, the connected correlator (see Fig.~\ref{loop&field} (left)) of a plaquette $U_{P}$ with two Polyakov loops is implemented on the lattice~\cite{DiGiacomo:1990hc}, as a natural modification of the connected correlator of a plaquette with a Wilson loop. The difference is that studies at finite temperature are made possible through the use of
\begin{equation}
\rho_{P}^{conn}=\frac{\left\langle \mathrm{tr}\left(P\left(x\right)LU_{P}L^{\dagger}\right)\mathrm{tr}P\left(y\right)\right\rangle }{\left\langle \mathrm{tr}\left(P\left(x\right)\right)\mathrm{tr}\left(P\left(y\right)\right)\right\rangle }-\frac{1}{N}\frac{\left\langle \mathrm{tr}\left(P\left(x\right)\right)\mathrm{tr}\left(P\left(y\right)\right)\mathrm{tr}\left(U_{P}\right)\right\rangle }{\left\langle \mathrm{tr}\left(P\left(x\right)\right)\mathrm{tr}\left(P\left(y\right)\right)\right\rangle },\label{eq:rhopconn}
\end{equation}
$N$ being the number of colors. The color field distribution is probed by varying position and orientation of $U_{P}$. In particular, $E_{l}\left(x_{t}\right)$ is measured with $U_{P}$ parallel to the plane formed by the Polyakov loops, at distance $x_t$ from it. The reason one uses $\rho_{P}^{conn}$ is that in the naive continuum limit 
\begin{equation}
 \rho_{P}^{conn}\xrightarrow[a\rightarrow0]{}ga^{2}\left[\left\langle F_{\mu\nu}\left(x\right)\right\rangle_{q\overline{q}}-\left\langle F_{\mu\nu}\left(x\right)\right\rangle_{0}\right],\qquad F_{\mu\nu}\left(x\right)=\sqrt{\frac{\beta}{2N}}\rho_{P}^{conn}\left(x\right).
\end{equation}

Measurements are accomplished in a multi-step procedure. For a given $\beta$, an ensemble of thermalized configurations, and then ensembles of ``smeared'' configurations after 6 to 20 APE smearing steps~\cite{11} (smearing factor $\alpha=0.5$) are generated. Then, $E_{l}\left(x_{t}\right)$, averaged over each smeared ensemble, is determined for different values of $x_{t}$, by means of~(\ref{eq:rhopconn}). Results are fitted with~(\ref{eq:fitfunct2}) (see Fig.~\ref{loop&field} (right)) and the parameters $\Phi$, $\mu$, $\nicefrac{\lambda}{\xi_{\nu}}$ and $\kappa$ are plotted versus the smearing step (Fig.~\ref{param} and Table~\ref{tab}): for each parameter a plateau is eventually searched in this plot. Smearing has been preferred to cooling, since it is safer in reducing fluctuations at finite temperature.

We remark that the present results are preliminary and refer to one $\beta$ value only.
After the possible optimization of smearing factor and distance between Polyakov loops,
other $\beta$ values must be considered, in the search for continuum scaling.
Assuming, however, that the continuum is reached at $\beta=6.1$ and a plateau is formed 
at about 16 smearing steps, we can estimate
\begin{equation}
 \nicefrac{\mu}{\sqrt{\sigma}}=3.137(299),\qquad\lambda=\nicefrac{1}{\mu}=0.150(14)\,\mathrm{fm},
\end{equation}
which agree, within error bars, with~\cite{5}, where the connected correlator of 
plaquette and Wilson loop was used, together with the cooling procedure to achieve noise reduction.

\begin{table}[tb]
\begin{center} 
\caption{Summary of the fit values for $SU(3)$, on a $20^{4}$ lattice, at $\beta=6.1$, statistics=1200.} \label{tab}
\small {\begin{tabular}{cccccc}
\hline\hline Smearing& $\phi$ & $\mu$ & $\lambda/\xi_\nu$ & $\kappa$ & $\chi_r^2$ \\ \hline6&4.478(0.602)&0.550(0.206)&0.740(0.680)&0.688(0.819)&0.17\\ 7&5.274(0.510)&0.523(0.146)&0.732(0.502)&0.678(0.603)&0.07\\ 8&5.792(0.404)&0.508(0.106)&0.712(0.369)&0.655(0.441)&0.03\\ 9&6.136(0.339)&0.499(0.086)&0.692(0.299)&0.630(0.354)&0.01\\ 10&6.357(0.307)&0.496(0.079)&0.661(0.264)&0.594(0.310)&0.01\\ 11&6.496(0.253)&0.495(0.066)&0.630(0.210)&0.558(0.244)&0.01\\ 12&6.588(0.224)&0.497(0.060)&0.599(0.181)&0.522(0.209)&0.02\\ 13&6.649(0.202)&0.498(0.056)&0.570(0.159)&0.489(0.181)&0.02\\ 14&6.689(0.184)&0.500(0.053)&0.544(0.142)&0.460(0.159)&0.03\\ 15&6.715(0.170)&0.502(0.050)&0.521(0.128)&0.434(0.142)&0.04\\ 16&6.733(0.158)&0.504(0.048)&0.500(0.117)&0.411(0.129)&0.04\\ 17&6.743(0.147)&0.505(0.046)&0.482(0.108)&0.390(0.117)&0.04\\ 18&6.749(0.139)&0.506(0.045)&0.465(0.100)&0.372(0.108)&0.04\\ 19&6.751(0.131)&0.507(0.044)&0.449(0.094)&0.355(0.100)&0.04\\ 20&6.751(0.125)&0.508(0.043)&0.435(0.089)&0.340(0.094)&0.04\\\hline\hline 
\end{tabular}}
\end{center}
\end{table}

\end{document}